\RequirePackage{fix-cm}
\documentclass[smallextended]{svjour3}       
\usepackage{graphicx}

\journalname{Transport in Porous Media}
\begin{document}

\title{Tomographic Study of Internal Erosion of Particle Flows in Porous Media
}
%


\author{Filippo Bianchi \and  Falk K. Wittel \and Marcel Thielmann \and Pavel Trtik \and Hans J. Herrmann}


\institute{Filippo Bianchi \at
              Computational Physics for Engineering Materials, DBauG, ETH Zurich, Stefano-Franscini-Platz 3; CH-8093 Zurich, Switzerland \\
              Tel.: +41-44-633 35 62\\
              \email{fbianchi@ifb.baug.ethz.ch}           
           \and
           Falk K. Wittel \at
              Computational Physics for Engineering Materials, DBauG, ETH Zurich, Stefano-Franscini-Platz 3; CH-8093 Zurich, Switzerland \\
              Tel.: +41-44-633 28 71\\
              \email{fwittel@ethz.ch}   
              \and
	         Marcel Thielmann \at
              Bayerisches Geoinstitut, University of Bayreuth, D-95440 Bayreuth, Germany \\
              Tel.: +49-921 55 3721\\
              \email{marcel.thielmann@uni-bayreuth.de}   
              \and
              Pavel Trtik \at
              Neutron Imaging and Activation Group, Laboratory for Neutron Scattering and Imaging, Paul Scherrer Institut, CH-5232 Villigen PSI, Switzerland \\
              Tel.: +41-56-310 5579\\
              \email{pavel.trtik@psi.ch}   
              \and  
              Hans J. Herrmann \at
              Computational Physics for Engineering Materials, DBauG, ETH Zurich, Stefano-Franscini-Platz 3; CH-8093 Zurich, Switzerland \\
              Tel.: +41-44-633 27 01\\
              \email{hans@ifb.baug.ethz.ch} 
}

\date{Received: date / Accepted: date}

\maketitle

\begin{abstract}
In particle-laden flows through porous media, porosity and permeability are significantly affected by the deposition and erosion of particles. Experiments show that the permeability evolution of a porous medium with respect to a particle suspension is not smooth, but rather exhibits significant jumps followed by longer periods of continuous permeability decrease. Their origin seems to be related to internal flow path reorganization by avalanches of deposited material due to erosion inside the porous medium. We apply neutron tomography to resolve the spatio-temporal evolution of the pore space during clogging and unclogging to prove the hypothesis of flow path reorganization behind the permeability jumps. This mechanistic understanding of clogging phenomena is relevant for a number of applications from oil production to filters or suffosion as the mechanisms behind sinkhole formation.
\end{abstract}

\section{Introduction}\label{intro}
Flow through porous media is concentrated in preferred channels that form a complex network \cite{Morais-etal-2009,Ebrahimi-etal-2014}. For particle-laden fluid flows at low Bagnold and Reynolds numbers distinct zones of particle deposition and erosion emerge \cite{herrmann-andrade_araujo-2007,araujo-andrade-2006,herrmann-andrade_etal-2006}. Hence the pore space is subject to continuous change. Larger zones of connected pores can be disconnected from the entire network if critical pores are clogged. On the other hand permeability jumps can occur, when critical pores are opened due to erosion. Fluid mechanics simulations recently revealed the mechanism behind the erosive bursts to be the pressure differences up- and down-stream from a clogged zone \cite{Jaeger-etal_2017}. The subsequent flow reorganization results in clogging of other channels and thus in a different channel network. Depending on the field of application, deposition or release of particles can have either positive or negative effects. Typical examples are erosion in filters \cite{Alem-etal-2015}, sand production in the petroleum industry - resulting in increased hardware abrasion and even clogging of pipes \cite{Acock2004,Dehghani2010} - or internal erosion like suffusion in geotechnical contexts \cite{Richards2007,Fannin2014,sherard-dunnigan-1989,frishfelds-etal-2011} resulting in failure of in embankment dams. Interestingly the same processes can be employed to engineer functionally graded fiber reinforced composites by infiltration with particle doped resins \cite{lundstroem-frishfelds-2012}. Changes in the pore geometry of packings due to colloidal deposition were studied \textit{ex-situ} by synchrotron X-ray difference micro-tomography \cite{Chen-etal-2008,Chen-etal-2009} with particle sizes of $\approx 1\mu$m in diameter for pure clogging. For larger particle sizes, however, \textit{ex-situ} studies are misleading due to particle deposition once the flow stops. In principle, particle flow in a porous medium can also be observed using index-matched materials for walls, particles and fluid \cite{Dijksman-etal-2012}, however for larger systems, subject to this study, even small mismatches in the refractive index result in indefinite images that prohibit further evaluation. Additionally, such experiments require a suspension of index-matched particles, which suffer from the same issues described above. To directly observe particle deposition and erosion, the particles in this suspension would also need to be dyed. In this case, the system would become opaque once a certain amount of particles have been deposited.

Recent developments in neutron tomography have made it possible to observe deposition and erosion on the pore scale \cite{Gruenzweig-etal-2013,Kaestner-etal_2016}. Compared to synchrotron X-ray tomography the acquisition times with neutron beams are longer due to lower fluxes. This is compensated however by the excellent fluid-matrix contrast arising from a high sensitivity to hydrogen \cite{Kaestner-etal_2016}. Apart from this, samples sizes used in this study are difficult to image at synchrotron sources, as they are too large for typical beams and exhibit unsuitable X-ray transmission for the energies usually available at synchrotron beam lines. An alternative imaging technique to neutron imaging (NI) for the study deposition of fines in through porous media could be based on magnetic resonance imaging (MRI) \cite{Sederman-gladden-2001,Britton-2007}. In a comparative study between NI and MRI, however for tomographic measurements NI showed superior resolution \cite{Oswald-etal-2015}. The used NI enables \textit{in-situ} studies of the process of channel reorganization in particle laden flows due to internal erosion with avalanches, resulting in clogging and unclogging in a realistic, three-dimensional pore scale.

Our main goal is to observe the reorganization of the pore space before and after permeability jumps that we observed in a previous study \cite{Bianchi-etal_2017}. We want to relate measured topological changes in pore space to observed permeability jumps, which are measured in parallel. This could lead to a new perspective on the flow dynamics of particle-laden flows with respect to today's picture.

The paper is organized as follows: First, the experimental setup is described in detail with focus on the test cell, the neutron beam line and the binary decomposition scheme of the image acquisition. Thereafter, we describe the applied image processing procedures, the pore space segmentation and pore network construction used for the subsequent analysis. In the results section, we first relate the pressure evolution to the one of the global attenuation. Then we look at the spatiotemporal evolution of the effective attenuation averaged horizontally and finally we resolve detailed changes in the pore network at distinct points in time.
\section{Experimental Setup}\label{setup}
\begin {figure}[htb]
\includegraphics[width=0.5\columnwidth]{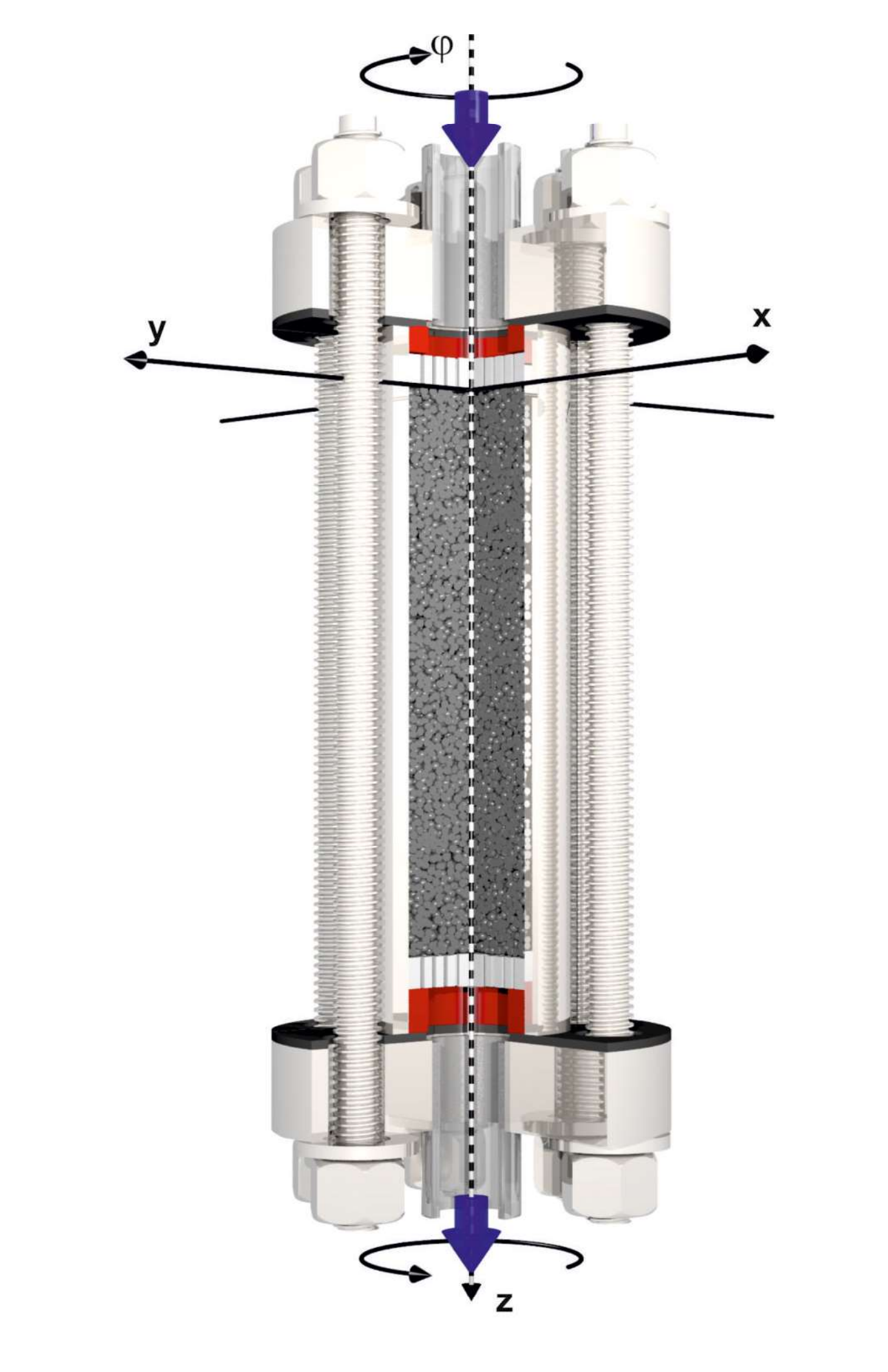}
\caption{\label{fig:testcell} Cut through the rotating test cell with definition of coordinates with segmented particle packing.}
\end{figure}

The porous medium in this study consists of a random closed packing of soda-lime glass beads with a diameter range of [0.9-1]~mm. A 16~mm wide cylindrical test volume of $\approx$80~mm height is filled and consolidated by vibration (150~Hz for 30~s). The packing is held in place by two 3D printed polycarbonate sieves with conical grooves of 0.4~mm width. This ensures homogeneous fluid penetration without clogging to minimize edge effects at the inlet and outlet. Compressive stresses are applied via a rubber to ensure the immobility of grains throughout the test (see Fig.~\ref{fig:testcell}). The test cell is mounted on a rotating stage that allows to take images under various angles $\phi$. Initially the pore space is filled with deionized boiled water and air voids are removed by flushing in both directions and shaking the sample. Note that entrained air bubbles can easily be identified and excluded from the tomographic data. This is the reference configuration at time $t_0$.

The experiment starts by continuously pumping an aqueous suspension with silica powder (230 mesh, Sigma Aldrich) of $d_{50}$=25~$\mu$m at a solid volume fraction $\Phi$ of 0.126 and 0.131 from a stirred surge tank through the porous medium (in $z$-direction, see Fig.~\ref{fig:testcell}). A controllable peristaltic pump (Gilson MINIPULS 3) is used, whose induced pressure oscillations are reduced by a damper. The use of a damper is necessary because otherwise pump-induced oscillations would dominate the temporal evolution of permeability, making its analysis difficult. The type of damper used (air chamber) has the effect of storing fluid when pressure rises, fluid which is eventually released when the pressure drops. The effective permeability during the experiment is obtained by tracking the differential pressure $\Delta p$ (Keller PRD-33 X with $\pm$145~Pa accuracy). The flux induced by the peristaltic pump is not perfectly constant and varies  depending on the experimental conditions. With increasing pressure, we observed a decrease in the flow rate. We therefore estimate the flux by counting the droplet rate of the flowing suspension at the filter outlet. This estimate results to be around 0.44 and 0.53~$\pm$0.15~ml/s. The Reynolds number of the flows in the experimental filters ranges between 1.5 and 1.8, while the Bagnold number is 0.4. In this setup a high-precision flow-meter was intentionally not used to avoid the exposition of the sensitive electronics to high energy radiation. Note that a full experimental campaign with focus on the effect of solid volume fractions and flux, including quantitative flux measurements was published in \cite{Bianchi-etal_2017}. With time, $\Delta p$ increases. For larger values of $\Phi$, the system can clog with and without jumps, resulting in strong pressure increase \cite{Bianchi-etal_2017}. The system is considered as clogged, when $\Delta p>3~$bar. To minimize the runtime of the experiment, we chose volume fractions which showed a clear tendency to clogging within several hours.

Spatio-temporal insight on flow inside porous media is rarely possible, but can be obtained non-invasively by neutron tomography \cite{Eyndhoven-etal-2015,Trtik-etal_2016}. In our case, however, we do not intent to directly resolve flow, but particle depositions and erosion, that occur on a larger time scale. As a neutron beam travels straight through a sample, its line integral represents the total attenuation $\Sigma$. The projection $i$ on a detector, acquired by a camera is called neutron radiograph  $I_i$ and a set of radiographs $I^K(t)={I_1,I_2,...I_m}$ at different rotation angles $\phi_i$ of the obstacle can be used to extract information about the spatial distribution of attenuation $\Sigma(x,y,z,t)$ by a reconstruction algorithm at a certain time $t$. Each material in the test cell has a characteristic attenuation coefficient. All metal parts are made of aluminum, since it is more or less transparent with respect to thermal neutrons ($\Sigma_{Al}=0.0105~$mm$^{-1}$). However water has a rather high attenuation coefficient  ($\Sigma_{H_2O}=0.565~$mm$^{-1}$) due to hydrogen nuclei, silica a much lower one ($\Sigma_{SiO_2}=0.0287~$mm$^{-1}$), and mixtures of those two materials exhibit an effective attenuation coefficient of 
\begin{equation}\label{eq:ruleofmix}
\Sigma=\Sigma_{H_2O}(1-\Phi)+\Sigma_{SiO_2}\Phi. 
\end{equation}
In other words, as a certain region of the pore space clogs, the silica concentration increases locally and consequently the effective attenuation in the clogged region decreases.

\begin {figure*}
\includegraphics[width=1\columnwidth]{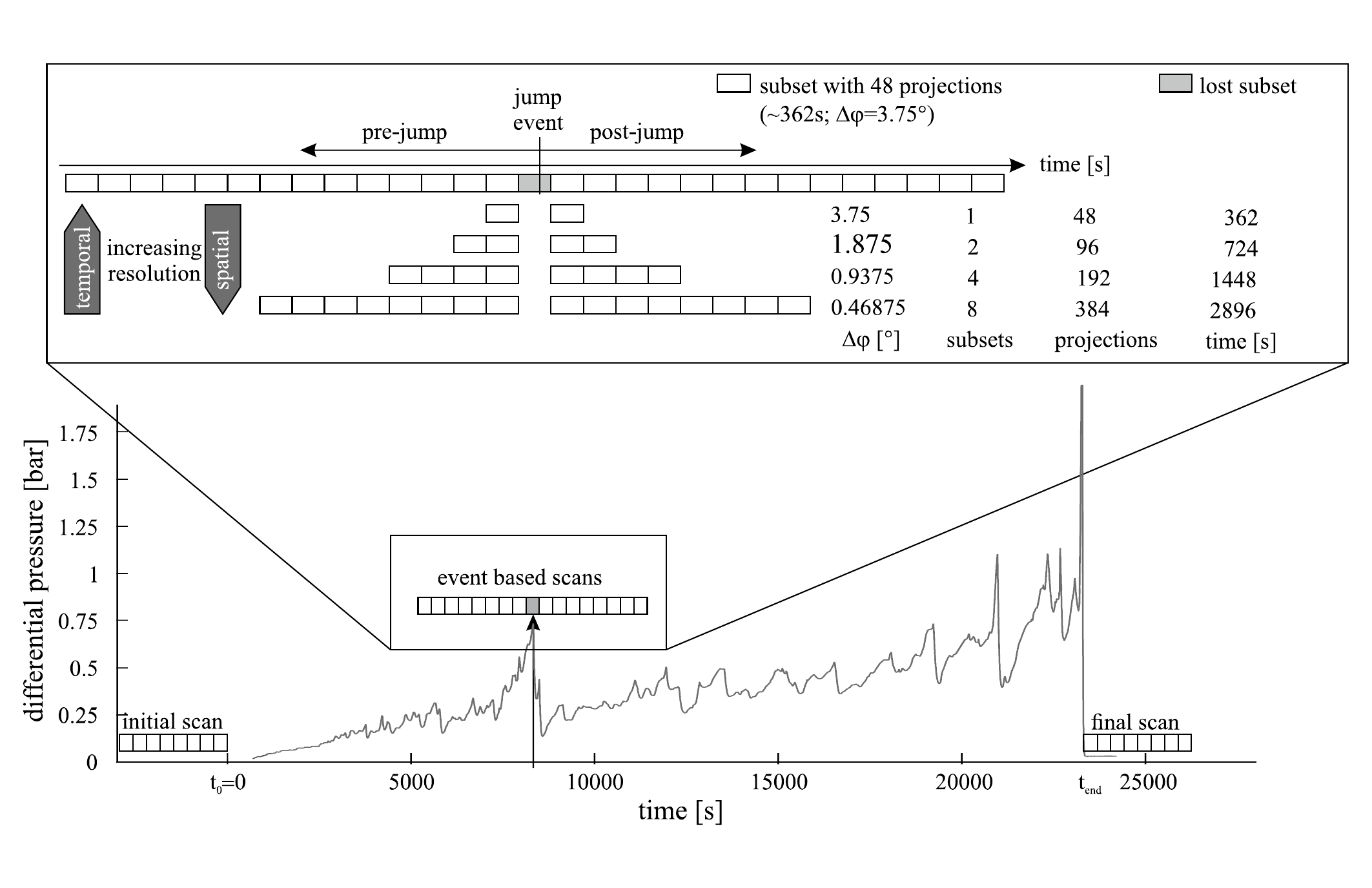}
\caption{\label{fig:decomposition} Image acquisition strategy with binary decomposition and differential pressure evolution of sample 2. The magnification shows the possible acquisition time windows before and after an event for 1, 2, 4, or 8 subsets.}
\end{figure*}

\begin {figure}[!hbtp]
\includegraphics[width=0.5\columnwidth]{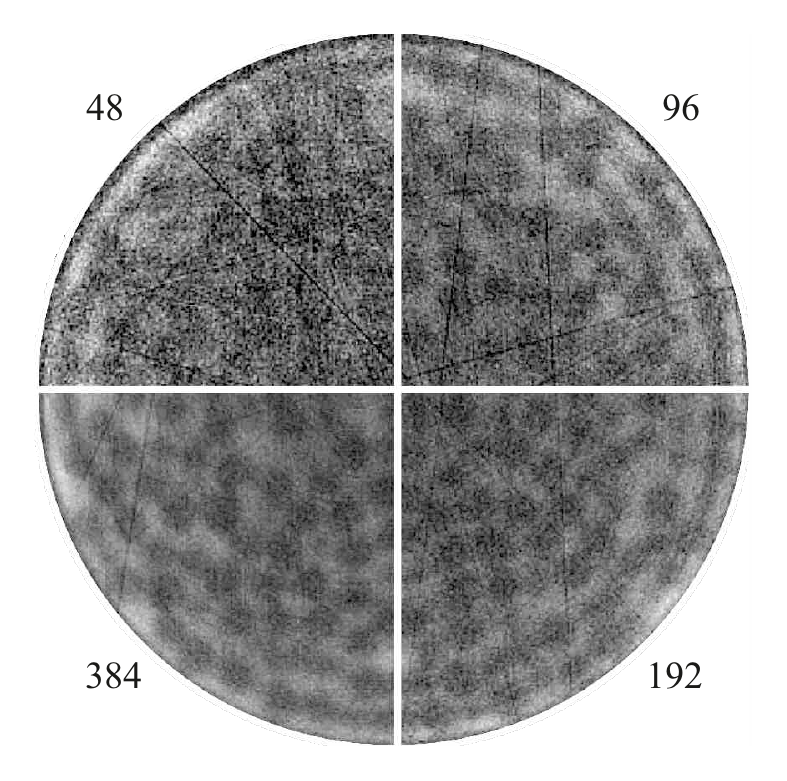}
\caption{\label{fig:subsets} Reconstruction levels with number of projections.}
\end{figure}

This work is based on experiments performed at the Swiss spallation neutron source SINQ, Paul Scherrer Institute, Villigen, Switzerland. The neutron beam is in the thermal spectrum i.e., neutrons with a most probable energy of 0.025~eV are produced. The facility features a high collimation ratio, low gamma background and a large beam diameter, described in detail in Ref. \cite{Lehmann-etal-2001}. Images were acquired by a scintillator-CCD camera-system with 2048$\times$2048 pixels resulting in a spatial resolution of 45~$\mu$m per pixel. The exposure time in the experiments was around 3~s per radiograph, depending on the beam intensity. Including the time for rotation of the stage and opening and closing of the shutter, the minimal time in-between radiographs is about 6.5~s. The quality of the tomographic reconstruction strongly depends on the number of projections. With a typical rotation angle increment of $\Delta\phi=0.5^{\circ}$ the total acquisition time of a scan sums up to $\approx 2900$~s. Since experiments cannot be paused and restarted without changes in the configuration due to rearrangements and deposition, tomographies have to be made simultaneously. For the reference scan and final scan after clogging, this is irrelevant and full scans were made. However, at intermediate times, larger changes in the solid fraction field during the acquisition of a scan, result in significant motion artifacts. To minimize this problem, a binary scan decomposition strategy is applied \cite{Kaestner-etal-2011}. The angle sequence of projections in the interval $[0,\pi)$ is chosen in such a way that subsets of subsequent projections - in this work 48$\rightarrow \Delta\phi=3.7^{\circ}$ - comprise sufficient information for reconstructions. Starting angles $\phi^j_0$ of the subsequent $j$=8 subsets follow the sequence $\Delta\phi\cdot[0,\frac{1}{2},\frac{1}{4},\frac{3}{4},\frac{1}{8},\frac{5}{8},\frac{3}{8},\frac{7}{8}]$. Therefore each subset, two, four or even eight consecutive or previous subsets can be used at any point in time for the reconstruction, depending on whether the intrinsic processes require a high temporal resolution or whether processes are slow enough to allow for a reconstruction with high spatial resolution (see Fig.~\ref{fig:decomposition}). Note that consequently the subset containing the jump event itself is lost for reconstruction.
Fig.~\ref{fig:subsets} shows the outcome of the reconstruction at different temporal resolution obtained by the inverse Radon transform using a filtered (Ram-Lak filter) back projection (FBP) algorithm \cite{ct-ref-88} implemented in Matlab2015. Prior to reconstruction, spot artifacts from gamma radiation are removed. Due to varying beam intensity, all projections had to be normalized by the current beam intensity, measured at image regions next to the projection of the test cell. Finally, all radiographs were corrected with flat and dark field projections. Note that the number of projections in a single subset with sparse image data of only 48 projections barely meets the Nyquist criterion \cite{ct-ref-88} for the reconstruction. Hence the minimal temporal resolution is $\approx$724~s per scan.
\section{Image processing and analysis}\label{sec:imageprocessing}
For the further analysis of the particle flow, it is essential to segment the solid skeleton formed by the grain packing from the reconstructed scans, and to extract the corresponding pore space. As grains are immobile throughout the experiment, it is sufficient to identify and segment them using the initial scan with the largest contrast in attenuation between silica and water. As the reconstructed data is noisy, blurred, and contains ring and star artifacts typical for FBP reconstructions, a meaningful segmentation requires a sequence of image restoration, enhancement and transformation and 3D filtering steps described in the following.

First the sample is aligned in the vertical direction by identifying the center of mass of slices, applying a linear interpolation along the $z-$axis and aligning the images in the $x-y-$plane. All pixels that are part of the cylindrical porous medium with radius of 180 pixels are now identified for further operations, while the others are ignored in the following. The intensity of all scans is now linearly stretched to the full 16 bit intensity range using the extremal intensities of the initial scan. The contrast is further stretched to the interval defined by the 10\% resp. 90\% percentile of all minimal, resp. maximal intensities of all slices of the initial scan. 

In the next step ring and star artifacts are reduced. First strong gradients are identified by taking the difference of each slice with its median filtered image \cite{Prell2009}. This image is binarized, and line shaped blops are identified, kept, and dilated, while others are removed. Finally for every voxel of a slice that is part of a blop, a new intensity in the scan is assigned by interpolating between the two nearest neighboring voxels in $z$-direction, that are not part of a blop.
\begin {figure}[!hbtp]
\includegraphics[width=.8\columnwidth]{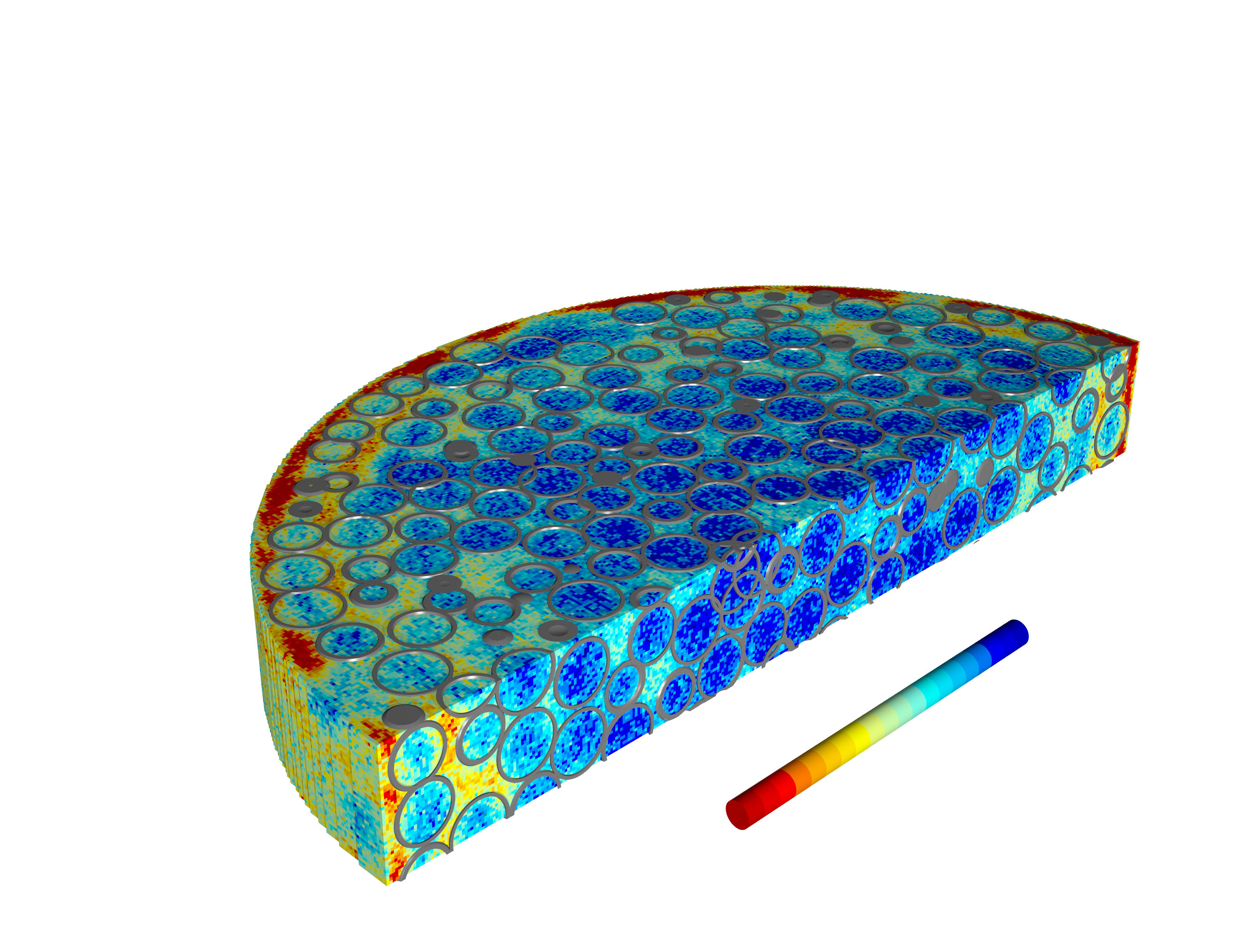}
\caption{\label{fig:particalid} Identification of particles by the attenuation coefficient field. Minimal to maximal values range from red to blue.}
\end{figure}

The solid skeleton is identified only on the reference scan at $t_0$. To segment grains successfully, first the scans need further image restoration such as compensation for nonuniform illumination using opened and closed scans, spatial filtering with 3D Gaussian kernels to eliminate noise, spatial convolution with 3D Laplacian kernels for deblurring and cropping regions that are not part of the region of interest (ROI). Thereafter, Otsu's method is applied for segmentation. Grains should be spherical objects with $\approx$22~voxels diameter and a volume of $V_g\approx$5500~voxels. However contact points avoid a clean segmentation of individual grains calling for morphological image processing. Already a large portion of grains can be isolated by morphological opening with spherical structuring elements. Blops with volume$> V_g$ are identified, and further morphological opening with increasing structuring sphere radius is applied, until the blop decomposes into its individual grains. Finally all blop and grain centers are calculated and stored. The identified solid grain packing with density 0.61 for sample 2 is rendered in Fig.\ref{fig:testcell}. From now on only the segmented pore space is further analyzed.

The solid phase can now be used to obtain the average intensity value of the silica and water respectively, used to map an intensity to an attenuation and finally local solid volume fraction $\phi$ using Eq.~\ref{eq:ruleofmix}. With these values, the intensity of pore space voxels of all reconstructed scans can be transformed into $\phi$ (see Fig.~\ref{fig:particalid}). Note that the evident cupping artifacts are irrelevant for difference images used in the analysis. Obviously, further image restoration is required to identify evolving structures inside the pore space. First all particle volumes and the cylinder walls are marked, so they can be excluded from the further evaluation. The remaining image only carries the information of the pore space and the noise is reduced by applying a 3D weighted Gaussian filter on all pore space voxels, where only pore space voxels enter. 

\begin {figure}[!hbtp]
\includegraphics[width=.8\columnwidth]{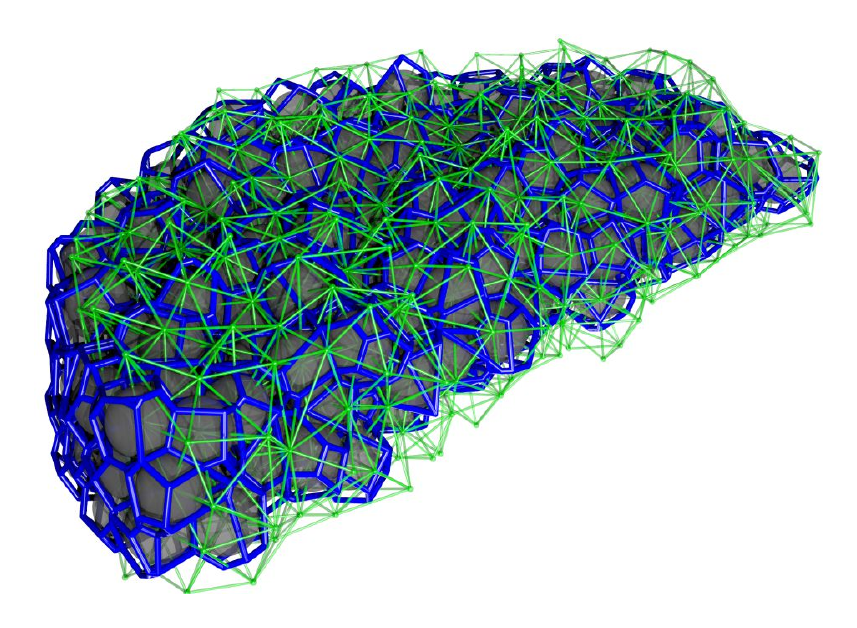}
\caption{\label{fig:evalproc} Construction of pore space based on Voronoi polyhedra (blue) via Delaunay triangulation of particle centers (green). }
\end{figure}

Once the solid skeleton is segmented with satisfying accuracy, the next step is to construct the pore space, based on tetrahedra and Voronoi polyhedra similar to \cite{Wang-etal-2012,Melnikov-etal_2015}. First, a layer of regularly spaced ghost particles is added on all surfaces. Then the centers of all identified particles and the ghost ones are triangulated. The Voronoi polyhedra are calculated from the triangulation. Each Voronoi polyhederon comprises one particle, while its vertexes are located on pores and edges ideally on pore channels (see Fig.~\ref{fig:evalproc}). In a small spherical sampling region around each vertex an average attenuation at the pore is obtained. Note that the deviation from the much more laborious identification of pore space voxels, located inside tetrahedra was found to be negligible. Consequently we do not distinguish between the particles concentration in the pore and the pore throat and assume the concentration to be homogeneous on each vertex. 
This way the attenuation field is reduced to the pore channel network, making it accessible for further evaluation.
\section{Results}\label{sec:results}
\subsection{Pressure evolution}\label{subsec:pressure}
During the experiments on average the differential pressure rises with time, until either the filter completely clogs or the pump is stopped. Pressure jumps are frequently observed.

\begin {figure}[!hbtp]
\includegraphics[width=1\columnwidth]{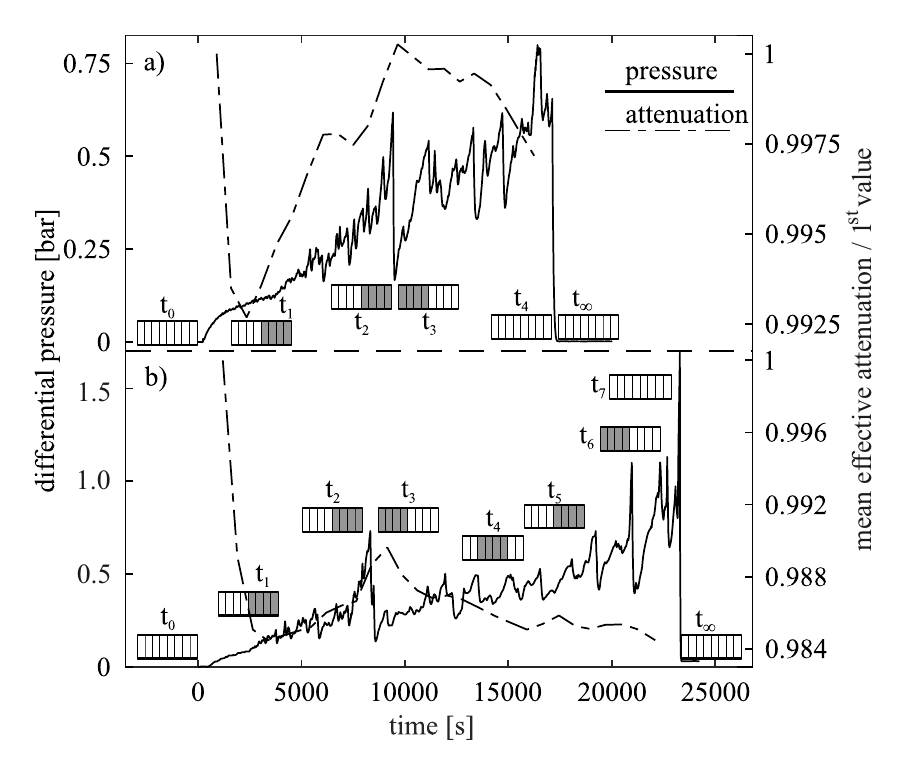}
\caption{\label{fig:PressureAttenuation} Evolution of differential pressure (solid line) and of mean effective attenuation (dash-dot line) during the experiments run with sample 1 (a) ($\Phi$~=~0.126) and sample 2 (b) ($\Phi$~=~0.131). The times at which the reconstructions are performed are depicted as t\textsubscript{0} to t\textsubscript{$\infty$}. The boxes mark evaluated acquisition intervals with 8 (entire) or 4 (dark) subsets each. The effective attenuation is normalized by the initial value, reconstructed from 192 projections each, resulting in an acquisition time window of 1448~s. Note that we represent the time in the middle of the acquisition time window after 96 projections.}
\end{figure}

The pressure evolution can be divided into three different phases (see Fig.~\ref{fig:PressureAttenuation}). During the first phase pressure increases smoothly without significant jumps. This phase lasts until the pressure reaches 0.1-0.2~bar, about 3000-5000~s. After this phase, jumps are observed. They occur repeatedly during the experiments. The last phase is characterized by the complete clogging of the sample (which occurs only in Fig.~\ref{fig:PressureAttenuation}b), when the pressure rises beyond the capacity of the pump. The pump is therefore stopped and the differential pressure drops to 0~bar.

\subsection{Evolution of effective attenuation}\label{subsec:attenuation}
The effective attenuation is the mean of all voxels of the pore space, either as a global mean value obtained during an acquisition period or height dependent. The global effective attenuation is shown for samples 1 and 2 in Fig.~\ref{fig:PressureAttenuation}. For comparison, it is normalized by the value at the first time window with 900s / 1200s for sample 1 / 2. Decreasing values represent increasing particle concentration or clogging, while for increasing ones particle concentrations decrease. Note that reconstructions from images acquired before the start $t_0$ or after the clogging $t_\infty$ cannot be included, since air bubbles in one case or particle sedimentation in the other result in significant attenuation shifts. After the start, in both samples first water is replaced by the suspension that continuously deposits, corresponding to the first phase of the pressure evolution with monotonic, jump-free increase. Surprisingly, the overall particle concentration does not increase, but decrease, as the system approaches the first larger pressure jumps. Hence one suspects a reorganization in form of a localization, where certain zones are clogging while in other regions the particle concentration decreases. The time of the pressure jumps and the sharp change from decreasing to increasing particle concentration coincide for both samples. What follows then is a more or less continuous increase of particle concentration until finally the maximum pressure is reached. 

\begin{figure}[hbtp]
\includegraphics[width=1\columnwidth]{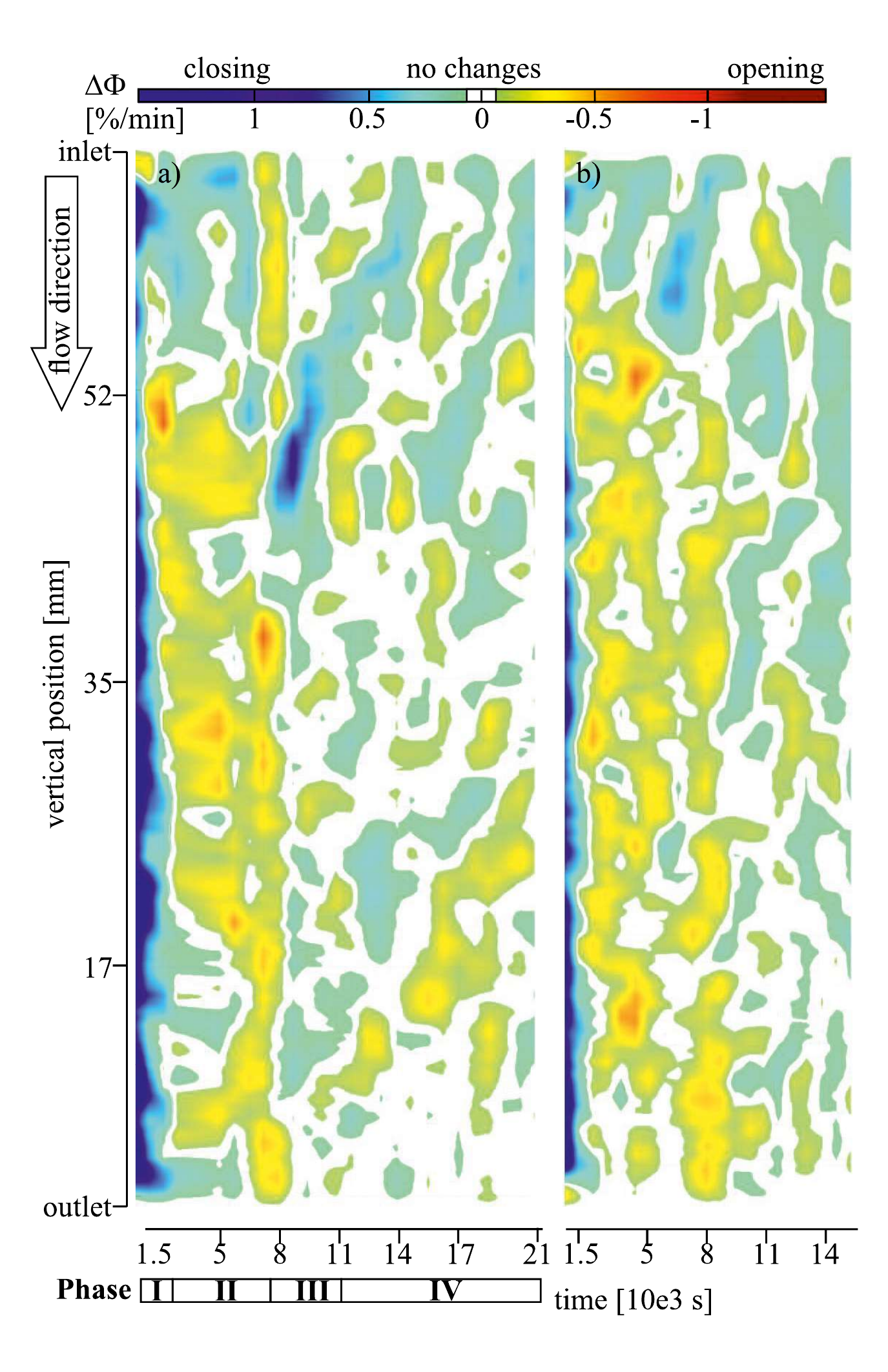}
\caption{\label{fig:attenuation_rates} Spatio-temporal evolution of the change rate of the averaged solid volume fraction in the pore space sample 2 (a) and 1 (b) reconstructed from 192 projections each, sampling a time window of 1448~s. The vertical axis gives the vertical position inside the cylinder in mm, the horizontal axis the time in s, while colors represent changes in the height dependent average of effective attenuation in intensity values.}
\end{figure}

When looking at the change rate of silica solid fraction or attenuation between reconstruction intervals over the height of the sample (see Fig \ref{fig:attenuation_rates}), rearrangements of silica become visible. Average effective attenuations are calculated by averaging attenuation values of all voxels of the pore space in a certain height. Using the attenuation values for pure silica and water from the initial scan at $t_0$ and Eq.~\ref{eq:ruleofmix}, attenuation values are translated to solid fraction values. One can observe the spatial evolution of deposition and erosion processes inside the filter. The evolution can be divided into four different phases, each one of them characterized by a particular behavior for sample 2 in Fig.~\ref{fig:attenuation_rates}a):
\begin{itemize}
\item Phase I (t $<$ 2000~s), increase of particle concentration along the whole filter;
\item Phase II (t = 2000 - 7600~s), closing of the top part and erosion in its wake;
\item Phase III (t = 7600 - 11000~s), erosion of the previously clogged region, which shifts downstream;
\item Phase IV (t $>$ 11000~s), several regions of clogging and erosion alternated in time and space.
\end{itemize}

During phase I, attenuation decreases with time over the whole sample height. This can be attributed to the gradual replacement of pure water by suspension, which therefore reduces attenuation. Additionally, suspended particles are deposited in a more or less similar fashion along the sample (see Fig.~\ref{fig:attenuation_rates}) which also contributes to a lower effective attenuation.

Phase II is characterized by an ongoing decrease in effective attenuation in the upper 13~mm of the sample (close to the inlet), while lower regions exhibit an increase in effective attenuation. The attenuation increase in the upper part is related to further particle deposition in this region, which thus acts as a filter. For this reason, less particles are transported downstream, which explains the increased attenuation in this region. Additionally, previously deposited particles can be washed out by the filtered fluid, thus adding to the increase in attenuation.

During phase III we observe an attenuation increase in the upper part of the filter which corresponds to particle erosion and pore opening. Right below this region (about 20~mm from the filter inlet), attenuation decreases at the same time, indicating particle transport from the filter in the adjacent regions downstream. Below this region with decreased attenuation, we observe another increase in attenuation, which can again be related to the washout of suspended and deposited particles. The shift of the low attenuation region downstream continues until 11000~s. Simultaneously, regions closer to the inlet also exhibit lower attenuation values, which indicates particle deposition in the upper part of the sample. This phenomenon can be most clearly observed at approximately 8000~s in sample 2 and corresponds to the first large pressure jump (see Fig. \ref{fig:PressureAttenuation}b)). 

After the first large erosion event, we observe in phase IV the formation of different regions that alternately feature attenuation increases or decreases. This shows that the pore network undergoes a highly dynamic process where deposition and erosion continuously alter spatial distribution. One recurring feature of this phase is the upstream migration of low attenuation regions. This indicates that particles are being deposited upstream of already clogged regions and that downstream erosion of those clogged regions could occur simultaneously. In the spatio-temporal evolution plot (Fig.~\ref{fig:attenuation_rates}) this is evident by a characteristic orientation of alternating erosion and deposition regions in both samples.

When comparing sample 1 and 2, similar features are visible. However, due to a lower solid fraction of the suspension, the processes appears to develop later and slower for sample 1, exhibiting different times for the phases. 
\subsection{Pore space observations}\label{subsec:porespace}
\begin{figure*}[hbtp]
\includegraphics[width=1\textwidth]{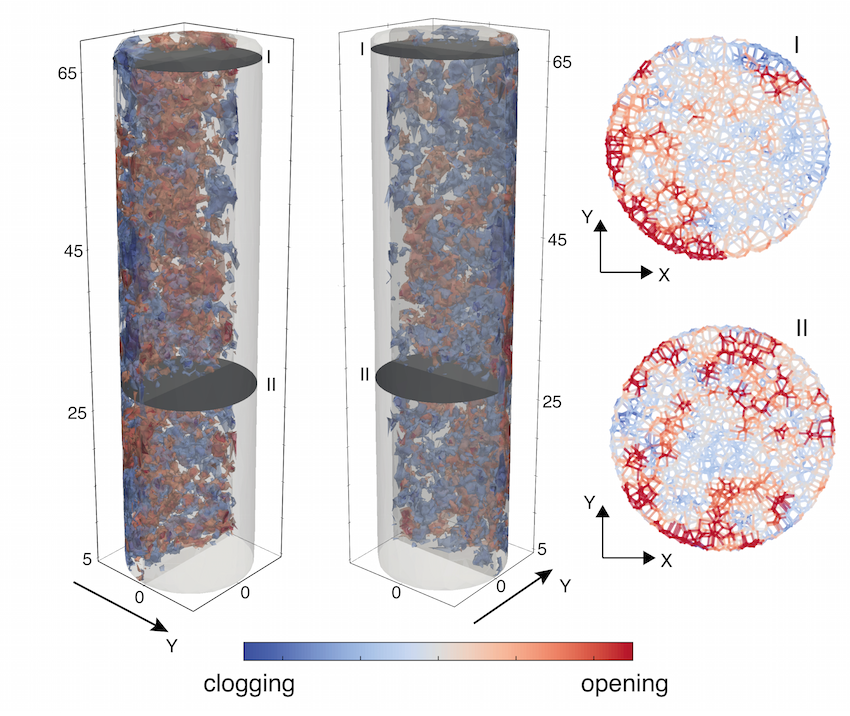}
\caption{\label{fig:attenuation_pores_t2_t3_all} Attenuation (and thus volume fraction) changes in the pore channels in sample 2 between $t_2$ and $t_3$. Red colors denote channels with increased attenuation, which corresponds to a decrease in silica content (opening). Blue colors denote channels with a decreased attenuation, which corresponds to an increase in silica content (clogging). The two plots on the left show the attenuation change in the whole sample, where the sample is sliced open along the x-z plane for better visibility. Length units are in mm. Grey circles denote the height at which horizontal profiles with a thickness of 2.26 mm are taken. The pore network of each slice (denoted by roman numbers) is shown in the right column.}
\end{figure*}

The the spatio-temporal evolution plot (Fig.~\ref{fig:attenuation_rates}) clearly shows that the porous medium constantly changes throughout the experiment. Note that preferential flow along the cell walls is not observed. To gain more insight in the sedimentation and erosion processes, we here have a closer look at the changes of the pore space of sample 2 between distinct jumps. To this end, we choose two jumps, namely the large jump between $t_2$ and $t_3$ and the jump between $t_5$ and $t_6$ . While the jump between $t_2$ and $t_3$ is clearly visible in in fig.~\ref{fig:attenuation_rates}, the jump between $t_5$ and $t_6$ cannot be clearly identified in the horizontally averaged silica volume fraction change. The silica volume change in the pore network between $t_2$ and $t_3$ is shown in fig.\ref{fig:attenuation_pores_t2_t3_all}. The two rows on the left side show the volume fraction change throughout the whole sample. The sample is sliced open along the x-z-axis for better visibility. The discrete pore network is extracted from two horizontal profiles located between 63 and 65.25~mm height (close to the inlet) and between 27 and 29.25~mm. Horizontal slices are denoted by grey circles with Roman numbers I and II and shown on the right side. 

During the jump between $t_2$ and $t_3$, we observe attenuation changes throughout the whole medium (as already observed in  fig.~\ref{fig:attenuation_rates}), with large regions experiencing attenuation increases and decreases. Close to the inlet (horizontal slice I), we observe that the silica content in the different pore channels is strongly heterogeneous, with different channel clusters having more and less silica content. In particular, a pore cluster close to the left lower boundary of the sample exhibits strong decreases in silica volume fraction. We thus presume that the opening of this cluster is responsible for the large pressure jump between $t_2$ and $t_3$. However, the occurrence of a jump does not imply a complete opening of the whole pore channel network. The simultaneous increase in silica content in the right side of the upper horizontal slice shows that the pore channel network is subject to significant topology changes during a jump and that even clogging of some regions might occur during this process.
In the lower horizontal slice, we observe the same behaviour, the only difference being that opening clusters are more distributed compared to the upper slice.

\begin{figure*}[hbtp]
\includegraphics[width=1\textwidth]{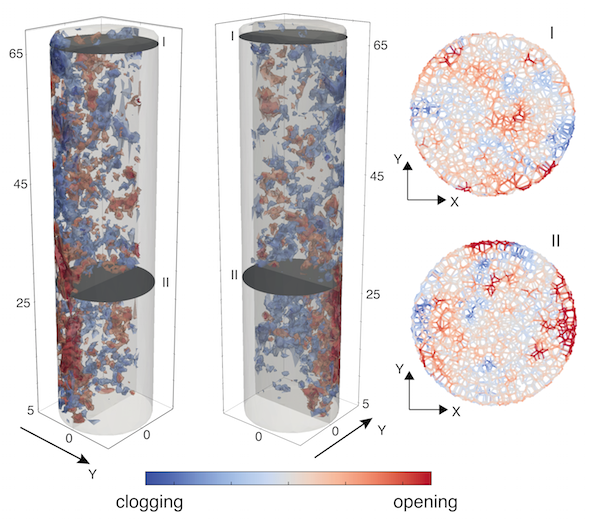}
\caption{\label{fig:attenuation_pores_t5_t6_all} Same as fig.\ref{fig:attenuation_pores_t2_t3_all}, but for the pressure jump between $t_5$ and $t_6$.}
\end{figure*}

When looking at the silica volume fraction change of the jump between $t_5$ and $t_6$ in figure \ref{fig:attenuation_pores_t5_t6_all}, a similar pattern can be observed. As could also be seen in fig.~\ref{fig:attenuation_rates}, the overall change in silica volume fraction is smaller compared to the jump between $t_2$ and $t_3$. Opening and clogging clusters are smaller and less connected in both the horizontal and vertical direction. In the upper horizontal slice, we again observe a cluster of pores that exhibits a strong decrease in silica content. The size of this cluster is much smaller than in the larger jump (which was to be expected). 
In the lower horizontal slice, we again observe several pore clusters with a significantly decreased silica volume fraction, with the largest of them extending over approximately the lower half of the sample height (see the two plots on the left side of fig.\ref{fig:attenuation_pores_t5_t6_all}.
\section{Discussion and Conclusions}\label{Sec:discuss}
We have studied the clogging mechanisms of deep bed filtration. We have shown that both deposition and detachment processes occur inside a filter. Zones of deposition and zones of resuspension appear and evolve with time. Clogging regions tend to move with time towards the sample inlet, unclogging regions are located in the wake of clogging regions. We have related large pressure loss jumps to general erosive events inside the filter. Such jumps are most likely related to i) the unclogging of pore clusters close to the inlet and ii) to the backward erosion in the downstream region of a clogged zone, resulting in resuspension of particles from the previously clogging zone. Observations of the pore space also show that the channel network is reorganized during such jumps, with the amount of rearrangement being proportional to the jump size.

In our measurements of the real 3D system, we did not observe the formation of a few discrete channels ( as in e.g. \cite{araujo-andrade-2006}), but only observed attenuation changes in larger zones. This might be due to the smaller dimensionality of the models presented in \cite{araujo-andrade-2006} or the fact that the flexibility of numerical models allows to approach a critical state more closely than what is possible with our experimental setup. Our results are in line with the results from \cite{Alem-etal-2015}, who investigated the clogging of sand columns and found that under constant flow conditions, the amount of straining (retention of small particles) is decreased compared to constant pressure conditions. In principle, from an algorithmic perspective more performant FBP algorithms \cite{Kaestner-etal_2016}, iterative reconstructions \cite{Eyndhoven-etal-2015} or wavelet based filters for ring artifact removal \cite{Munch:09} could be used. In a new campaign the temporal resolution could also be increased following ideas of Ref.~\cite{Trtik-etal_2016}. All such improvements would give a sharpened view on the system, increase the accuracy of the pore space segmentation, pore network construction and dynamics. However, the overall conclusions would not change.

It is not yet clear how the results of the present study can be upscaled to larger systems, where the number of channels is significantly increased and thus the clogging and reopening of single channels might have a much smaller effect on overall dynamics. Numerical simulations of the processes observed in this study (e.g. \cite{Jaeger-etal_2017}) are up to date only feasible on smaller systems and thus cannot be used for this purpose. A parameterized description of the deposition/erosion processes is thus needed, but these upscaling procedures always include the risk of missing important micro structural processes. The data set obtained with the experiments presented here could serve as an additional basis for further upscaling efforts, such as the one presented in \cite{frishfelds-etal-2011}. From an experimental view, the investigation of larger systems is possible, but comes at the cost of losing the possibility to monitor the pore-scale deposition/erosion processes. However, it might be possible that the onset of erosional events could be related to an increased noise level in pressure, flow and effluent concentration signals and possibly also acoustic emissions. The feasibility of such measurements will have to be investigated in the future.

As our model setup does not allow for the total erosion of the porous medium, a direct comparison to the issue of sand production is difficult. \cite{Fannin2014} used the terms „suffusion“ and „suffosion“  to distinguish between non-destructive and the destructive erosion processes. Nevertheless, we expect the avalanche-like resuspension events we observe in our experiments to be equally - if not more - probable in boreholes. When making analogies to sanding and backward erosion, it is important to keep in mind, that in our experiments the pore space, determined by the silica grains, does not change. Nevertheless our findings are relevant for several filtration-related industrial applications. The evolution of effective attenuation and of the channel network indicates that clogging is not a homogeneous process along the whole filter, providing information for an improvement of filter design and efficiency.

\begin{acknowledgements}
The neutron imaging leading to the results was obtained under project proposal no. 20150745 at the NEUTRA beamline at PSI / SINQ. We acknowledge the excellent support of Peter Vontobel and Eberhard Lehmann. The authors are grateful for the financial support of the ETH Zurich under Grant No. 06 11-1, and the European Research Council (ERC) Advanced Grant 319968-FlowCCS. A full dataset used in this work can be found on researchgate.net under the creative commons license.
\end{acknowledgements}

\bibliographystyle{spmpsci}      
\bibliography{allcitations}   

\end{document}